\documentstyle[11pt,epsfig,here]{article}

\def\m0{\hbox{$m_{\hbox{\scriptsize 0}}$}}
\def\mg{\hbox{$m_{\hbox{\scriptsize 1/2}}$}}
\def\A0{\hbox{$A_{\hbox{\scriptsize 0}}$}}
\def\Atau{\hbox{$A_\tau$}}
\def\tb{\hbox{$\tan\beta$}}
\def\Gcs{\hbox{GeV/$c^2$}}
\def\Tcs{\hbox{TeV/$c^2$}}
\def\ecs{\hbox{eV/$c^2$}}
\def\mchpm{\hbox{$m_{\chi^\pm}$}}
\def\msnu{\hbox{$m_{\tilde\nu}$}}
\def\mchi{\hbox{$m_\chi$}}
\def\mgravc{\hbox{$m_{\hbox{\scriptsize $\tilde G$}}^2$}}
\def\mgrav{\hbox{$m_{\hbox{\scriptsize $\tilde G$}}$}}
\def\mslr{\hbox{$m_{\tilde \ell_R}$}}
\def\mser{\hbox{$m_{\tilde e_R}$}}
\def\mstr{\hbox{$m_{\tilde\tau_R}$}}
\def\mstau{\hbox{$m_{\tilde\tau}$}}
\def\mtau{\hbox{$m_{\tau}$}}
\def\lplm{\hbox{$\ell^+\ell^-$}}
\def\epemto{\mbox{$\hbox{e}^+\hbox{e}^- \to$}}
\def\epem{\mbox{$\hbox{e}^+\hbox{e}^-$}}
\def\PRep#1#2#3{{\it Phys. Rep. }{\bf #1 }(#2) #3}
\def\PRD#1#2#3{{\it Phys. Rev. }{\bf D#1 }(#2) #3}

\begin{document}
\vspace*{-1.8cm}
\begin{flushright}
\flushright{\bf LAL 99-58}\\
\vspace*{-0.5cm}
\flushright{October 1999}
\end{flushright}
\vskip 2.5 cm

\centerline {\LARGE\bf Interpretation of {\sc Higgs} and {\sc Susy}} 
\vspace{2mm}
\centerline {\LARGE\bf searches in {\sc
mSugra} and {\sc Gmsb} models}

\vskip 1.5 cm

\begin{center}
{\Large  \bf Jean Baptiste de Vivie}\\
E-mail:
devivie@lal.in2p3.fr
\end{center}

\begin{center}
{\bf\Large Laboratoire de l'Acc\'el\'erateur Lin\'eaire}\\
{IN2P3-CNRS et Universit\'e de Paris-Sud, BP 34, F-91898 Orsay Cedex}
\end{center}

\begin{abstract}
{\sc Higgs} and {\sc Susy} searches performed by the ALEPH
Experiment at LEP are interpreted in the framework of two constrained
R-parity conserving models: Minimal Supergravity
and minimal Gauge Mediated Supersymmetry Breaking.
\end{abstract}

\section{Introduction}
Searches for supersymmetric and Higgs particles have been 
performed u\-sing the data collected with the ALEPH detector at LEP, 
at centre-of-mass energies up to 189~GeV. All have given negative 
results. This can be interpreted as constraints on the parameter 
space of a given model.
The results are here presented in the framework of two different
R-parity conserving constrained models: Minimal Supergravity 
({\sc mSugra})~\cite{nilles} and Gauge Mediated Supersymmetry Breaking
({\sc Gmsb})~\cite{ambr} models. Scans of their parameter space are carried
out~\cite{papiermssm,gmsbscan} in order to set a lower limit 
on the mass of the lightest neutralino in {\sc mSugra} and on
the universal {\sc Susy} mass scale in {\sc Gmsb}. The detailed
interpretation of searches in such constrained frameworks is of
particular relevance to understand the interplay of the various
experimental constraints. Moreover, it allows a direct
comparison with TEVATRON limits.

\section{Interpretation in the Minimal Supergravity model}
\subsection{Model parameters and constraints}
The {\sc mSugra} model allows to reduce the parameter space of the Minimal
Supersymmetric Standard Model ({\sc Mssm}) down to four parameters and
one sign. In this framework, several constraints are assumed at the
Grand Unification (GUT) scale: all scalar soft masses (including those
pertaining to the Higgs sector) are equal to \m0, all gaugino masses are equal to \mg, and 
the trilinear couplings are unified in a single parameter \A0. These
three parameters are used as inputs to the renormalization group
equations, to determine the low energy parameters. The 
bilinear coupling between the two Higgs fields is traded for the ratio of their 
vacuum expectation values \tb. Finally, $\mu$ is determined up to a
sign ambiguity by requiring that Electroweak Symmetry Breaking be
dynamically triggered by radiative corrections due to the large top
Yukawa coupling.

From a given (\m0,\mg,\A0,\tb,sign($\mu$)) set, all couplings and
masses of {\sc Susy} particles are computed. Several constraints
are applied to define the allowed parameter sets. First, theoretical constraints
are imposed to determine whether a set is physically acceptable: 
no particle should be tachyonic, a proper Electroweak
Symmetry Breaking is required and the top Yukawa coupling should not develop a
Landau pole up to the GUT scale. The top quark mass is set to 175\
\Gcs. Experimental constraints are then applied. LEP1 limits on
non-standard contributions to the total or invisible Z width are
used. Nonetheless, the main r\^ole is played by direct searches 
for new particles at LEP2. Low \mg\ values are excluded by chargino 
searches and further constrained by slepton searches at low \m0.
Higgs boson searches allow for a wide exclusion in the low \tb\
regime. Finally searches for heavy stable charged particles are 
also used to deal with quasi-stable staus. 

\subsection{Results in the (\m0,\mg) plane}
The results of the scan are first presented for $\A0 = 0$. Exclusion
domains in the (\m0,\mg) plane for $\tb
= 3$ (top) and 10 (bottom) are shown in Fig.~\ref{plan}. 
Chargino searches lead to a lower limit on \mg\
approximately independent of \m0, except for low \m0, where the experimental
sensitivity is reduced. In this region, slepton searches can be used. For $\tb =
3$ and negative $\mu$ where stop mixing is small, Higgs searches
exclude a large fraction of the parameter space.

The impacts of the further {\sc mSugra} constraints are
studied with a special attention payed to the {\it corridor} region, 
corresponding to small sneutrino masses. In this
region, the chargino production cross section is reduced due to
the negative interference between s and t-channel processes and two body decays of
charginos to lepton-sneutrino dominate, leading to almost invisible final states when $0 <
\mchpm - \msnu < 3$~\Gcs. 
An example of such a region is shown in
Fig.~\ref{mchilim}(a), for $\tb = 4.2$ and $\mu < 0$. The combination of the
Z width and the chargino constraints yields a lower limit on \mg\ of
63~\Gcs; adding slepton searches increases this limit to 94~\Gcs. Finally
Higgs searches allow to set a lower limit on \mg\ of 107~\Gcs,
corresponding to a $\chi$ mass of 45.4~\Gcs.

\subsection{Mass limit for the lightest neutralino}

\begin{figure}
\begin{center}
\epsfig{file=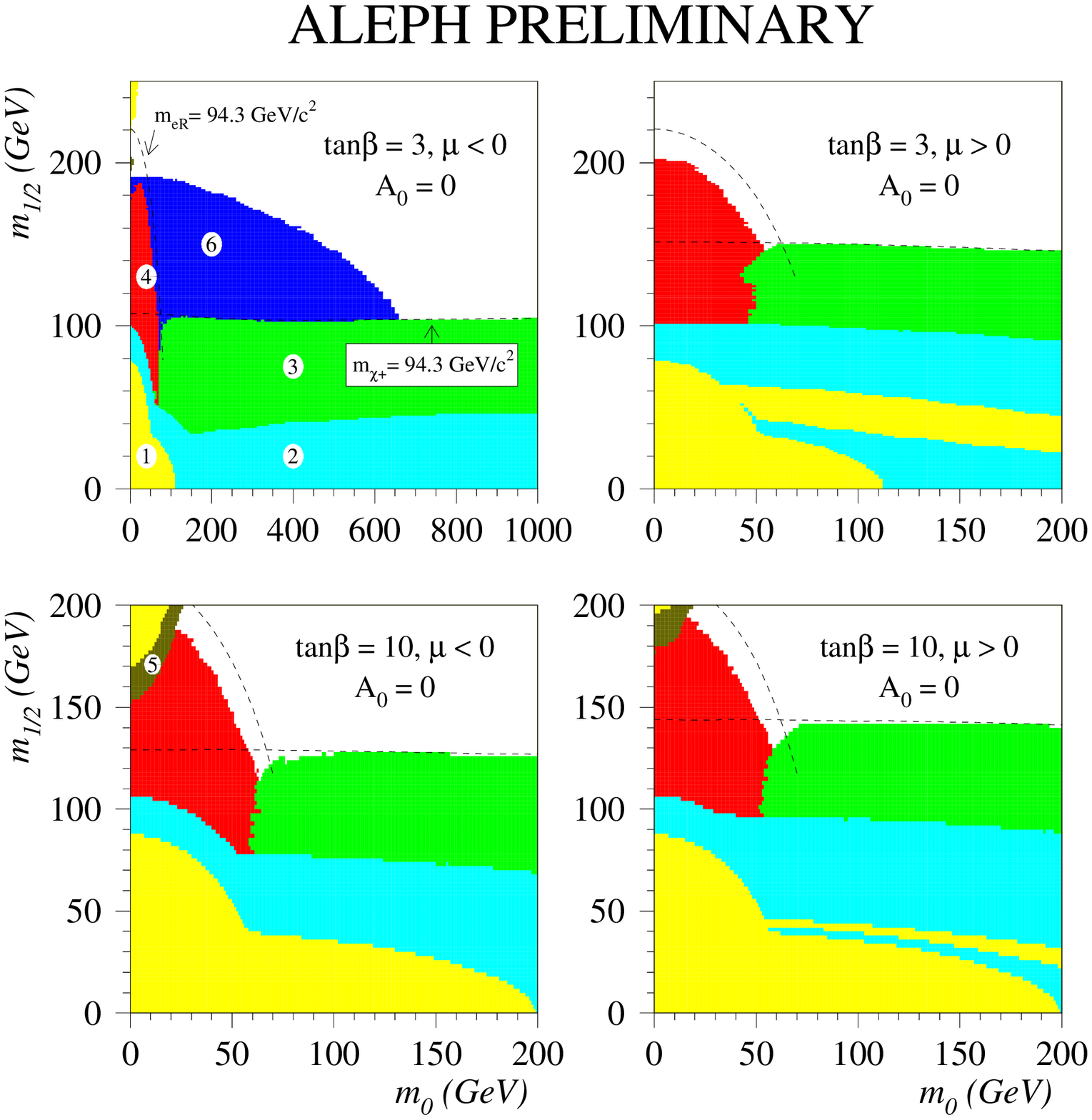,width=12cm}
\caption{\small Regions of the (\m0,\mg) plane excluded for $\A0 =
0$ and $\tb = 3$ (upper plots) and $\tb = 10$ (lower plots). 
Region 1 is theoretically forbidden. The other regions are
excluded by LEP1 constraints (2), chargino (3), slepton (4), 
heavy stable charged particle (5) and Higgs (6) searches. The thin dashed lines represent the kinematic limit for direct chargino and selectron searches.}
\label{plan}
\end{center}

\begin{center}
\epsfig{file=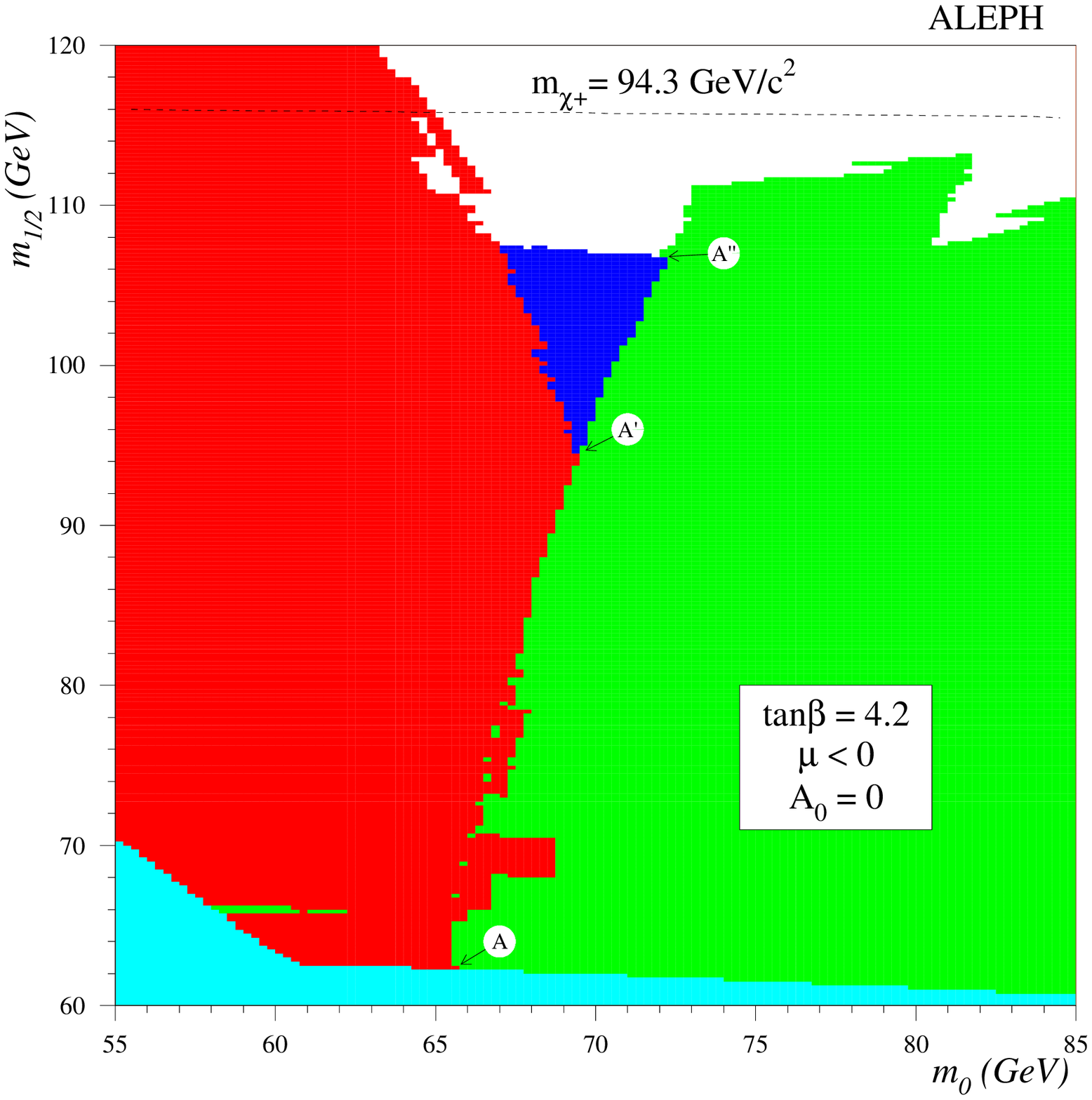,width=7cm}
\epsfig{file=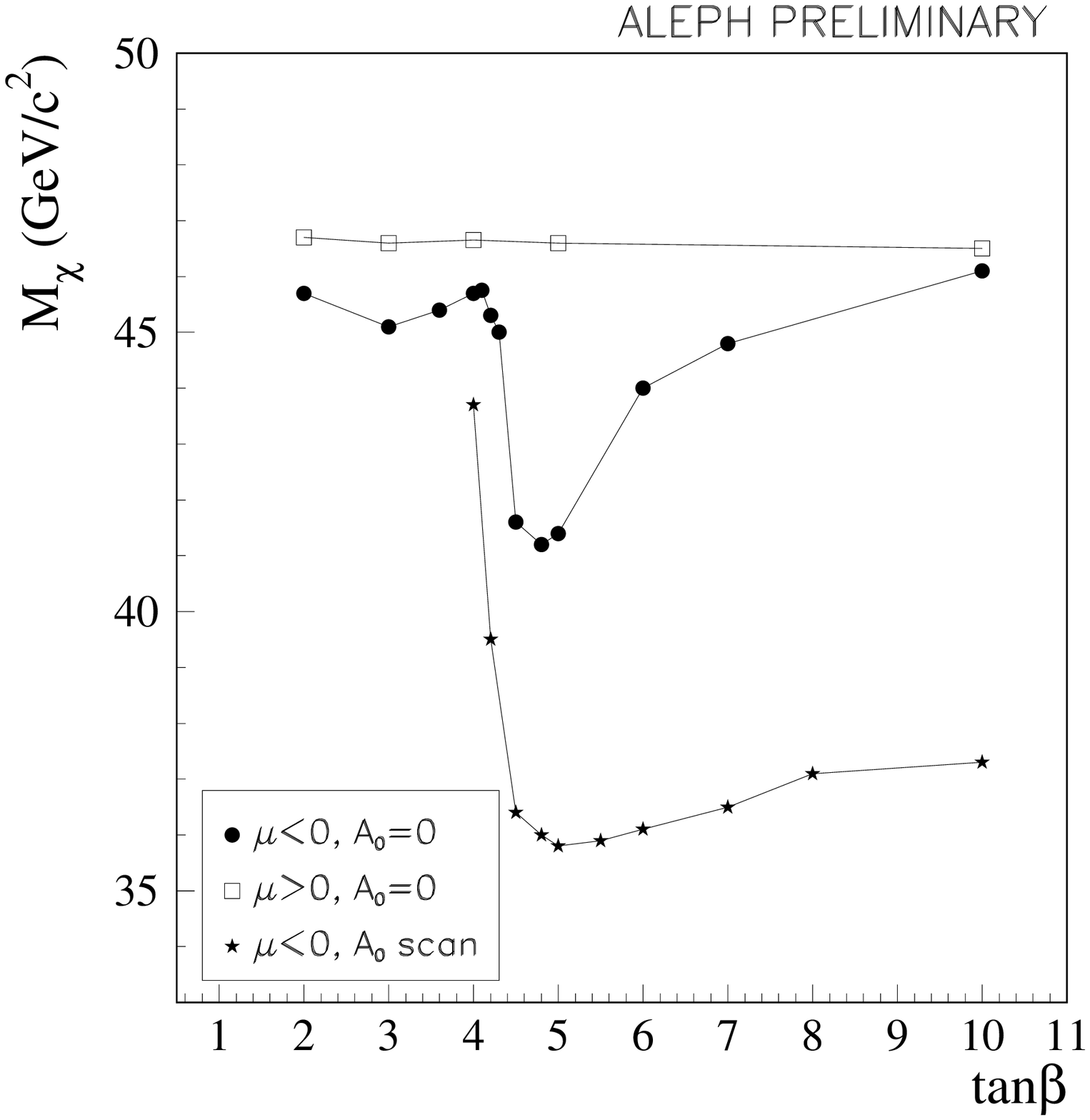,width=7cm}
\caption{\small Zoom of the corridor region for $\tb = 4.2$ and $\mu < 0$ (a).
Mass limit for the lightest neutralino as a function of \tb\ (b).}
\label{mchilim} 
\end{center}
\end{figure}

The mass limit for the lightest neutralino is presented in
Fig.~\ref{mchilim}(b), as a function of \tb. For positive $\mu$, it is
approximately independent of \tb. For negative $\mu$, a structure is
observed. At small \tb, Higgs searches are used to cover the 
corridor region. As \tb\ increases, Higgs searches become less
efficient and slepton searches become constraining. However, they
cannot completely cover the corridor for intermediate \tb, and
therefore a gap appears in the \mchi\ limit. Altogether, for $\A0 =
0$, the lower limit on \mchi\ is 41.5~\Gcs. 
However, as can be seen in Fig.~\ref{mchilim}(b), the impact of non 
zero \A0\ value cannot be neglected. Scanning over \A0\ leads to 
the final mass lower limit for the lightest neutralino within {\sc mSugra}, 
for $\tb \le 10$:
$$\mchi > 35.8\ \Gcs$$

This result improves by $\sim 3\ \Gcs$ on the one obtained in the 
less constrained minimal model (MSSM) considered in Ref.~\cite{papiermssm}.
The limitation to relatively low \tb\ values comes from the fact 
that the combined analyses are optimized for the case of negligible 
stau mixing. In the MSSM interpretation, the latter condition 
is met by fine-tuning the trilinear coupling \Atau. However, in 
{\sc mSugra} such an adjustment is not any longer possible; the
regions in the parameter  
space where the stau is the NLSP and nearly degenerate with the 
lightest neutralino are less covered by direct 
selectron searches. For these parameter sets, charginos
predominantly decay into $\tilde \tau \nu_{\tau}$ with very soft tracks from 
subsequent stau decays which are difficult to detect. In such cases,
the lower limit on \mg\ is 
set by the Z width measurement leading to a
degradation of the absolute lower limit on the LSP mass.  
To recover sensitivity in these cases, specific neutralino 
analyses should be included.

\vspace{-.4cm}
  
\section{New topologies in Gauge Mediated Supersymmetry Breaking models}
\vspace{-.2cm}
\subsection{Model parameters}
\label{mod}
\vspace{-.2cm}
The other generic class of models studied by the LEP collaborations
are the {\sc Gmsb} models. In this framework, Supersymmetry breaking
is mediated to the observable sector by a number of messenger pairs
charged under the Standard Model (SM) gauge groups. In minimal models,
five parameters and one sign are needed to determine masses and couplings: F, the
{\sc Susy} breaking scale which fixes the coupling of superparticles
to the gravitino, \tb, sign($\mu$), N, the effective number of messenger pairs, $\Lambda$,
the universal mass scale of {\sc Susy} particles and $\hbox{M}_{mess}$, the mean
messenger mass, used as starting point for the RGEs. In these models,
the soft masses are computed from N, $\Lambda$ and $\hbox{M}_{mess}$,
as well as the trilinear couplings which however are expected to be small.

\subsection{Relevant topologies}
In {\sc Gmsb}, the Lightest Supersymmetric Particle (LSP) is the
gravitino, which can be considered as massless and not interacting 
from an experimental point of view. The experimental topologies depend
on the nature of the Next to Lightest Supersymmetric Particle (NLSP)
and its lifetime, determined by the gravitino mass: $\tau \propto
\mgravc \propto (\hbox{F}/\hbox{M}_{pl})^2$. The lightest neutralino
and the sleptons are in general the only superparticles relevant for
LEP2 searches, in addition to the gravitino.\\

\vspace{-0.3cm}
{\it The lightest neutralino NLSP case}\\
\vspace{-0.2cm}
\\
In this case the relevant process is the pair-production of
neutralinos, decaying into a photon and a gravitino. For short
neutralino lifetimes, the experimental topology consists of two
acoplanar photons and missing ener\-gy. No excess of candidates has been  
observed in the data.


In the minimal model, and for very short lifetimes, this search
excludes $\chi$ masses smaller than 90~\Gcs.
Since the pair production of bino neutralinos 
can only occur via a t-channel selectron exchange, this limit depends slightly
on the selectron mass as shown in Fig.~3. This search allows to almost 
completely rule out the {\sc Gmsb} interpretation of the CDF event 
in the selectron pair-production scenario. 

For moderate lifetimes, the
previous search is complemented by a search for a single photon with
large ``impact parameter'' (non pointing photon), which takes advantage of the high
granularity of the ALEPH electromagnetic calorimeter. 
For very long lifetimes, neutralinos decay outside the detector, giving
rise to an invisible final state. \\

{\it The slepton NLSP case}\\
\\
Depending on \tb, two distinct {\it scenarii} can be considered: the
slepton co-NLSP scenario at low \tb, where all right handed sleptons
are mass-degenerate, and the stau NLSP scenario, at large \tb\ where
the stau becomes lighter due to a large mixing.
Again, topologies depend on lifetime. For
very short lifetimes, the standard {\sc Mssm} slepton searches for very
high $\Delta m$ (the mass difference between the slepton and the LSP) can be
used. The topology consists of a pair of acoplanar leptons and mis\-sing
energy. For intermediate lifetimes, sleptons can fly and decay in the
tracking volume and therefore give rise to detectable kinks. If they
decay inside the beam pipe, the associated track has a large impact
parameter. Finally, for long lifetimes, sleptons decay outside the
detector. The final state therefore consists of a pair of heavy stable
charged particles, with two main characteristics: the kinematics of the
pair-production and the high specific ionization in the Time
Projection Chamber. The number of candidate events observed in the
data and their properties are compatible with the
SM expectation. The negative results of these searches are 
translated into a lower limit for the mass of the slepton NLSP,
independent of its lifetime: in the co-NLSP scenario, $\mslr >
85~\Gcs$ and for the stau NLSP, $\mstr > 68~\Gcs$. 

To extend the sensitivity of the search in the case of short lifetimes, a
search for sleptons produced in cascade decays of neutralinos has been
performed. The process of 
interest is $\epemto \chi\chi \to \ell \tilde \ell \ell \tilde \ell \to
\lplm \tilde G \lplm \tilde G$. Here advantage is taken of the multi
lepton topology. Two leptons may be soft (from the neutralino decays)
and two are hard (from the slepton decays). Due to the majorana nature
of the neutralino, the charge (and flavour) of the two hard leptons
are uncorrelated, leading to spectacular final states, {\it e.g.} with 
two energetic equally charged muons. Moreover, the
neutralino production (via t-channel selectron exchange) 
cross section is large, since the neutralino is
mainly bino and the right handed selectron is light. For $\mchi < 87~\Gcs$ and
$\mchi - \mstau > \mtau$, this search is used to improve the stau mass
limit to 84~\Gcs.

\section{Scan of the {\sc Gmsb} parameter space}
\subsection{Ranges for the parameters and constraints}
In the absence of any excess with respect to the SM, all these
searches are used to put constraints on the parameter space. The five
parameters presented in~\ref{mod} are varied in order to cover a large
fraction of the parameter space, relevant for LEP2 studies and
compatible with the main motivations for {\sc Gmsb} models: five
values of N are probed (from 1 to 5), $\sqrt{\hbox{F}}$ is varied from
10\ to $10^4$ \Tcs, $\hbox{M}_{mess}$ from 10 to $10^9$ \Tcs,
$\Lambda$ from 1\ \Tcs\ to min($\sqrt{\hbox{F}}$,$\hbox{M}_{mess}$)
and \tb\ is required to be above 1.3. Several constraints are applied
on each parameter set. Direct searches for acoplanar taus and
heavy stable charged particles performed at LEP1 and constraints from
the Z width are used to exclude very low NLSP masses. Standard {\sc
Susy} searches for charginos and sleptons at LEP2 are also used to
cover long neutralino lifetimes. Finally all searches presented above
contribute to exclude a large fraction of the parameter space.

\subsection{Interplay between the various searches and mass limit for
the NLSP}

\begin{figure}
\vspace{-1cm}
\begin{center}
\epsfig{file=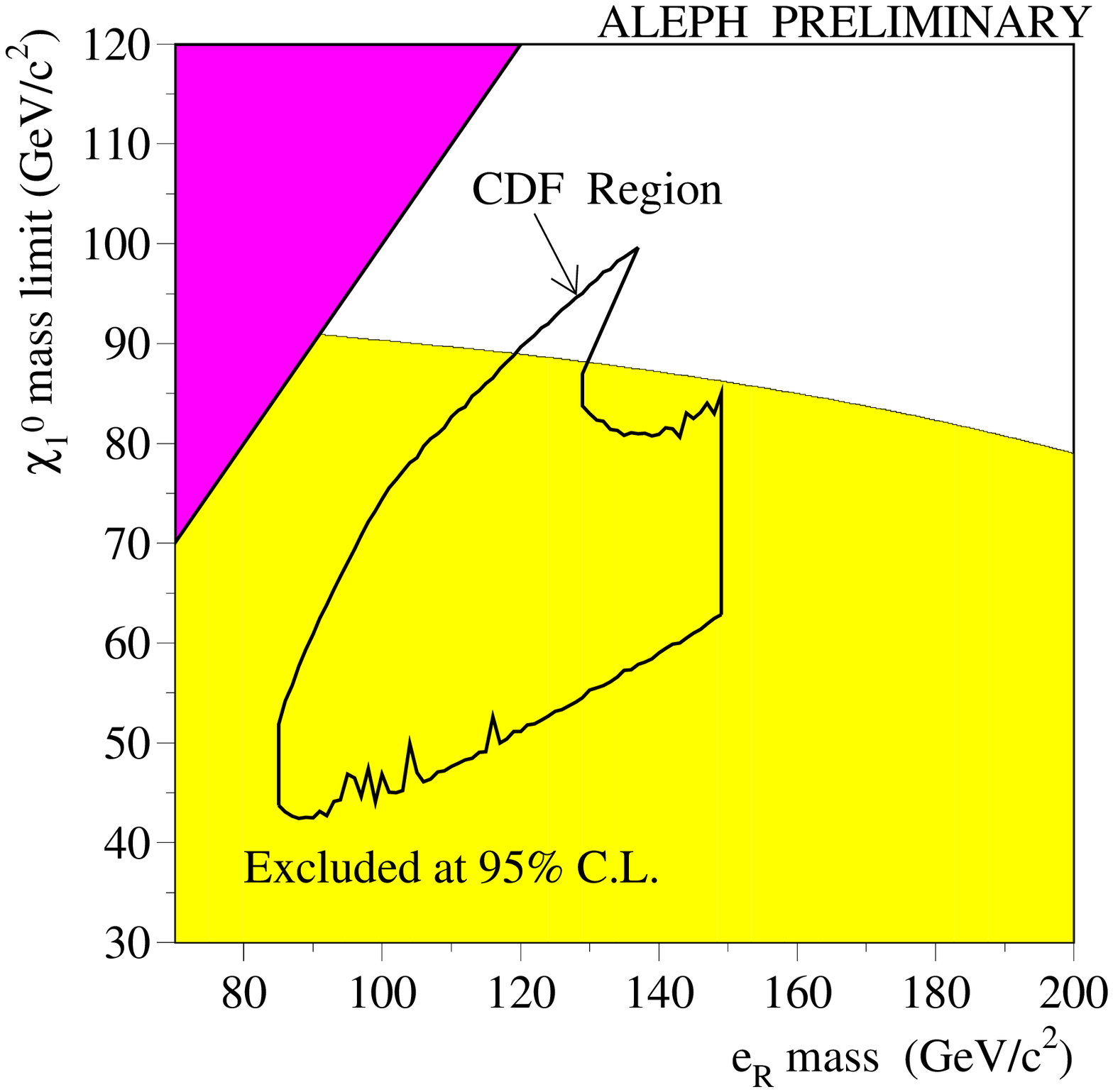,width=10cm}
\vspace{-0.6cm}
\caption{\small  Excluded region 
in the (\mser,\mchi) plane for a bino neutralino.}
\end{center}
\label{2phot} 

\begin{center}
\vspace{-0.5cm}
\epsfig{file=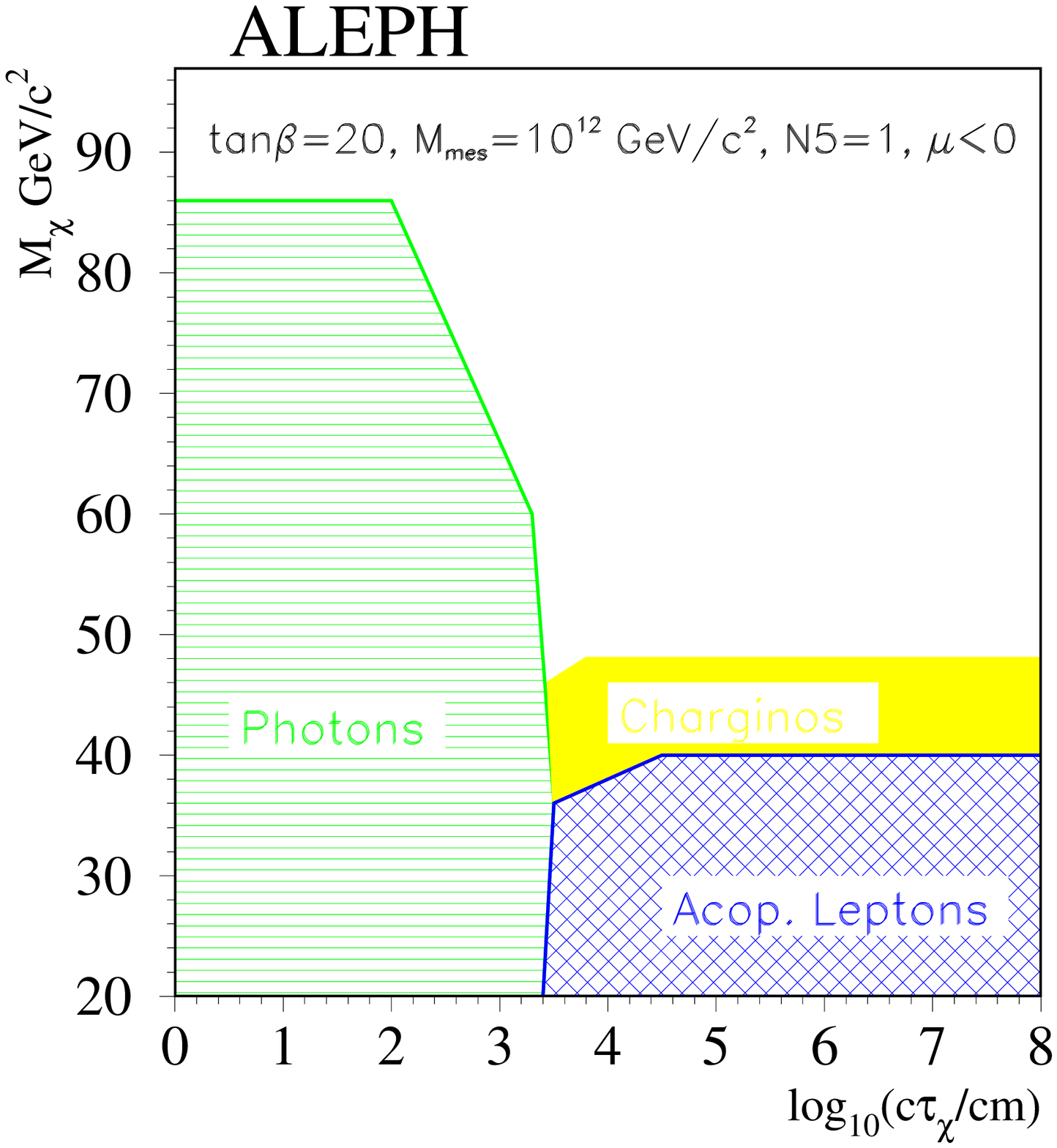,width=8cm}
\caption{\small Mass limit for the lightest neutralino as a function of its
lifetime.}
\label{comp}
\end{center}
\vspace{-0.7cm}
\end{figure}

An example of the complementarity of the standard {\sc Mssm} and {\sc
Gmsb} searches is shown in Fig~\ref{comp}, where N, $\hbox{M}_{mess}$
and \tb\ have been chosen in order for the lightest neutralino to be the
NLSP. For long lifetimes, the neutralino escapes the detector and
therefore only indirect limits can be set, as in the usual {\sc
Mssm}. For short lifetimes, the acoplanar photons complemented by the 
non pointing photon searches determine the limit. Finally in this
case, the lower limit on \mchi\ is 45~\Gcs.

The mass limit for the NLSP is obtained by varying all the parameters
in the considered range. The results of this scan, presented as
excluded regions in the (\mchi,\mstau) plane, are illustrated in
Fig.~\ref{nlspres}. It can be emphasized that all searches contribute, in
particular that the multi lepton topology   
significantly extends the exclusion domain for sleptons nearly
degenerate with the lightest neutralino. Altogether, the mass limit is
45~\Gcs\ for the neutralino NLSP, and 67~\Gcs\ for the stau NLSP.


\subsection{Lower limit on $\Lambda$}
\begin{figure}[h]
\vspace{-0.5cm}
\begin{center}
\epsfig{file=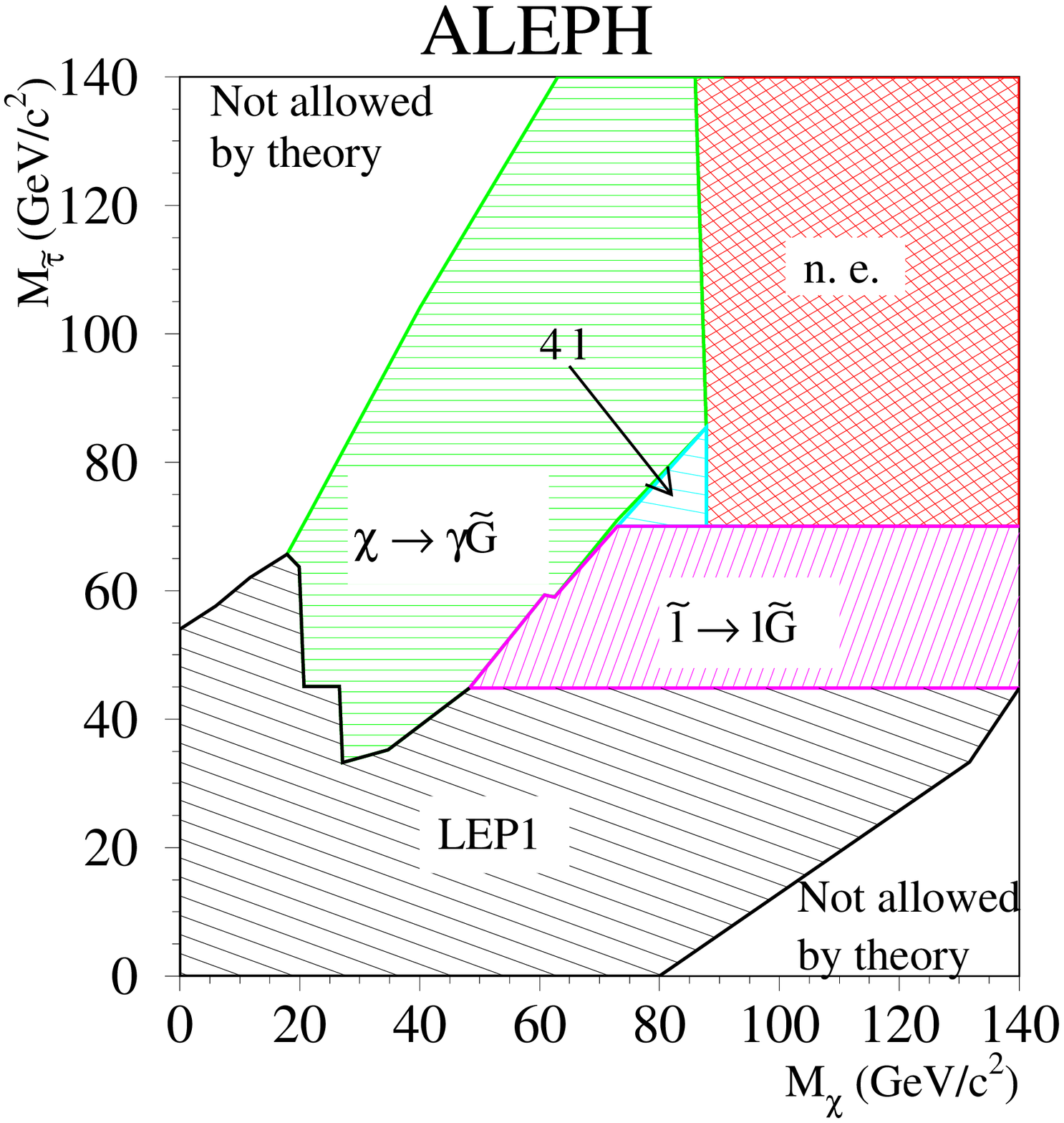,width=7.7cm}
\epsfig{file=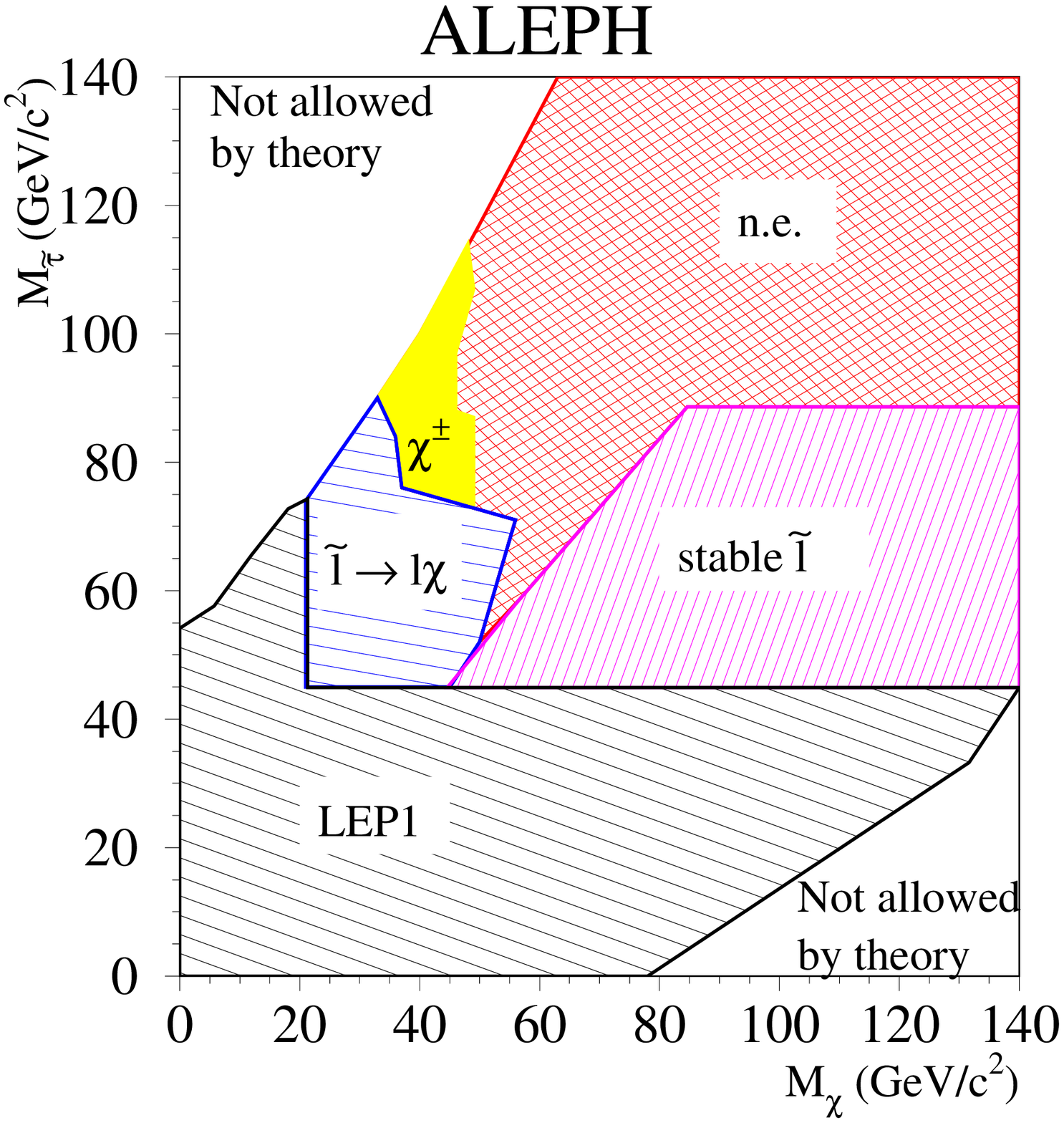,width=7.7cm}
\caption{\small Excluded regions in the
(\mchi,\mstau) plane for short (left) and long (right) NLSP
lifetimes. The {\it n.e.} region is not excluded and the
{\it not allowed by theory} regions correspond to regions
unaccessible within the parameter ranges considered.}
\label{nlspres}
\vspace{-.2cm}
\end{center}
\begin{center}
\epsfig{file=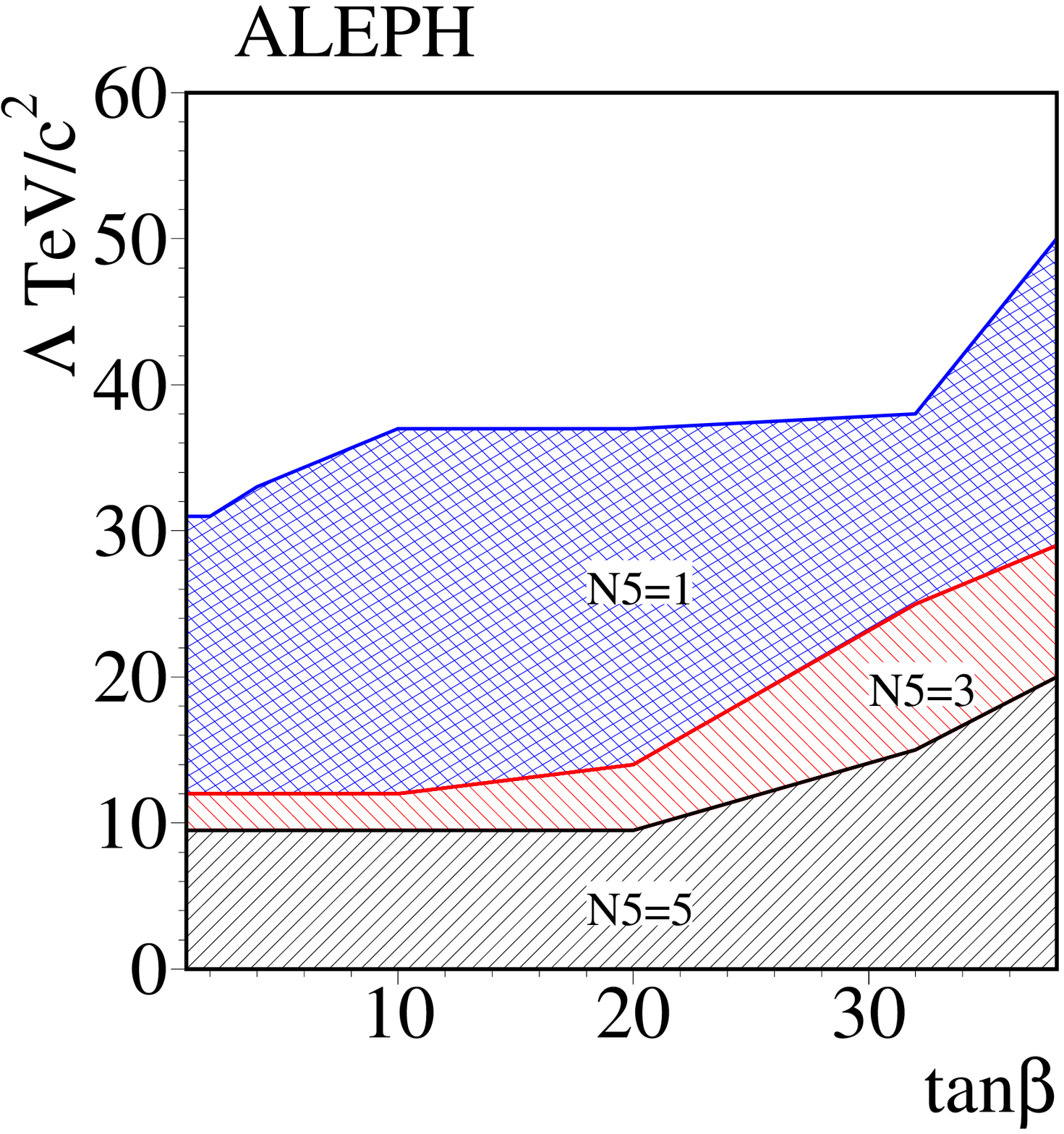,width=8cm}
\vspace{-0.5cm}
\caption{\small Lower limit on $\Lambda$ as a function of \tb\ for three
different values of the effective number of messenger pairs.}
\label{lam}
\end{center}
\end{figure}

In {\sc Gmsb} models, the parameter which sets the overall scale
for {\sc Susy} masses is $\Lambda$. The lower limit on $\Lambda$ as a function
of \tb\ obtained in this study is shown in Fig.~\ref{lam}. It is
essentially derived from the mass limit for the NLSP. Within the 
considered parameter ranges, the absolute lower limit on
$\Lambda$ is 9 \Tcs, found at low \tb\ for a quasi stable $\chi$ NLSP. This can be
translated into an indirect mass lower limit for the gravitino: $\mgrav~>~
2\ 10^{-2}~\ecs$.
\def \textfraction{0}
\def \textfraction{0.2}

\section{Conclusion}
The search for Supersymmetry is one of the paramount physics goal of
LEP2. A detailed interpretation of searches has been performed by the ALEPH
Collaboration in the constrained framework of {\sc mSugra}. For $\tb
\le 10$, this results in a lower limit on the mass of the lightest
neutralino: $\mchi>35.8\ \Gcs$. Nevertheless, there are regions in the
parameter space at large \tb\ where the constraints are weak and this limit 
is degraded. New specific topologies 
pertaining to {\sc Gmsb} models have been extensively searched for. 
In the absence of any hints for a signal, limits are set on masses and
model parameters. In particular, the overall mass scale of {\sc Susy} 
particles has to be larger than 9~\Tcs, almost independently of all
the other {\sc Gmsb} parameters.

\section*{Acknowledgements}
I would like to thank J.-F. Grivaz and M. Kado for their help in preparing this
talk. It is a pleasure to thank L. Duflot, G. Ganis and M. Kado for
careful reading and comments on this manuscript.


\begin{thebibliography}{99}
\bibitem{nilles} H. P. Nilles, {\em ``Supersymmetry, Supergravity and Particle Physics''},
\PRep{110}{1984}{1}.
\bibitem{ambr} S. Ambrosanio {\it et al.}, {\em ``Signals for
gauge-mediated supersymmetry breaking models the at CERN LEP2
collider''}, \PRD{56}{1997}{1761}.
\bibitem{papiermssm} The ALEPH Collaboration, {\em ``Search for Charginos and
Neutralinos in \epem\ Collisions at Centre-of-Mass Energies near
183 GeV and Constraints on the MSSM Parameter Space''},
CERN-EP/99-014, submitted to {\it European Physics Journal}.\\
The ALEPH Collaboration, {\em ``Search for Charginos and
Neutralinos in \epem\ Collisions at $\sqrt{\hbox{s}} =$ 188.6 GeV and Mass
Limit for the Lightest Neutralino''}, ALEPH 99-011, CONF 99-006.
\bibitem{gmsbscan} The ALEPH Collaboration, {\em ``Search for Gauge
Mediated {\sc Susy} Breaking topologies at $\sqrt{\hbox{s}} \sim 189$
GeV''}, ALEPH 99-045,CONF 99-024. 
\end{thebibliography}
\end{document}